\documentclass[aps,preprint,nofootinbib]{revtex4-1}

\usepackage{graphicx}
\usepackage{dcolumn}
\usepackage{bm}

\begin{document}


\title{Weighing in on the Higgs} 



\author{Jens Erler}
\email{erler@fisica.unam.mx}
\affiliation{School of Natural Sciences, Institute for Advanced Study,
Einstein Drive, Princeton, NJ 08540, USA}
\altaffiliation[Permanent address: ]{Departamento de F\'isica Te\'orica, Instituto de F\'isica,
Universidad Nacional Aut\'onoma de M\'exico, 04510 M\'exico D.F., M\'exico}

\date{\today}

\begin{abstract}
Assuming the validity of the Standard Model, or more generally that possible physics beyond it
would have only small effects on production cross sections, branching ratios and electroweak
radiative corrections, I determine the mass of the Higgs boson to $M_H = 124.5 \pm 0.8$~GeV
at the 68\% CL.  This is arrived at by combining electroweak precision data with the results of
Higgs boson searches at LEP~2, the Tevatron, and the LHC, as of december of 2011.
The statistical interpretation of the method does not require a look-elsewhere effect correction.
The method is then applied to the data available at the time of the 2012 summer conferences.
In this case, a remarkable bell-shaped $M_H$ distribution is observed, 
and $M_H = 125.5 \pm 0.5$~GeV is extracted. 
The significance of the bulk (signal) region of the distribution of neither experiment actually exceeds 
five standard deviations, but the combination implies a $6.8~\sigma$ effect. 
\end{abstract}


\maketitle 

\section{Introduction}
\label{intro}

The LHC Collaborations ATLAS~\cite{ATLAS} and CMS~\cite{CMS} have presented preliminary 
combinations of their Standard Model (SM) Higgs boson searches in data sets which correspond in 
the most sensitive channels 
to integrated luminosities of 4.6 to 4.9~fb$^{-1}$ of $pp$ collisions at $\sqrt{s} = 7$~TeV.
In addition, the CDF and D\O\ Collaborations combined results on searches in $p\bar{p}$ collisions 
at the Tevatron~\cite{CDFandD0:2011aa} at $\sqrt{s}=1.96$~TeV, based on luminosities ranging 
from 4.0 to 8.6~fb$^{-1}$.
In this brief communication, I analyze their findings simultaneously with earlier results from 
LEP~2~\cite{Barate:2003sz} and with constraints from electroweak precision data.
The goal is to obtain the most likely values for the mass, $M_H$, of the SM Higgs boson by taking 
all experimental information at face value and by explicitly accounting for any tensions or 
inconsistencies between data and background hypotheses, data and signal hypotheses, 
as well as (implicitly) between different data sets.
The statistical interpretation is unambiguous within Bayesian data analysis, 
the natural framework~\cite{Erler:2000cr,Degrassi:2001tg} for parameter estimation.  
(For an alternative approach, see Ref.~\cite{Flacher:2008zq}.)
Thus, I give an answer to the question: 
``Assuming the approximate validity of the SM and allowing all experimental information, 
what is $M_H$?".

This article is organized as follows: 
Sec.~\ref{data} describes both the method and the data used in this work.
The results are presented in Sec.~\ref{results}, while Sec.~\ref{outlook} gives conclusions and 
an outlook. 
Finally, Appendix~\ref{updates} incorporates significant experimental updates including 
first results from the LHC operating at $\sqrt{s} = 8$~TeV.

\section{Data}
\label{data}
\subsection{Data treatment}

The master equation used for this is given by,
\begin{equation}
p(M_H) = e^{-\chi^2_{\rm EW}(M_H)/2}\ Q_{\rm LEP}\ Q_{\rm Tevatron}\ Q_{\rm LHC}\ M_H^{-1},
\label{pmh}
\end{equation}
where the first factor is from the precision data, and 
$Q_{\rm LEP}(M_H)$ and $Q_{\rm Tevatron}(M_H)$ are the ratios of the likelihood for the signal of 
a particular $M_H$ hypothesis plus the background (H+B) to that of the background (B) 
alone~\cite{Erler:2010wa}.
Similarly, $Q_{\rm LHC} = Q_{\rm ATLAS}(M_H) \ Q_{\rm CMS}(M_H)$.
Unfortunately, for the latest ATLAS and CMS data these quantities have not been made publicly 
available.
I construct $Q_{\rm ATLAS}$ and $Q_{\rm CMS}$ through the relation,
\begin{equation}
2 \ln Q \equiv \chi^2_{H+B} - \chi^2_{B} \equiv 
\left( {1 - \bar\sigma_{\rm obs}\over \Delta\bar\sigma_+} \right)^2 - 
        \left( {\bar\sigma_{\rm obs} \over \Delta\bar\sigma_-} \right)^2,
        \label{qdef}
\end{equation}
where $\bar\sigma_{\rm obs}$ can be thought of as an effective observed cross section combining
the various channels considered by the LHC Collaborations.
It is normalized to the corresponding Higgs boson cross-section at the reference $M_H$, 
{\em i.e.\/} $\bar\sigma_S (M_H) \equiv 1$.
The errors are in general asymmetric, 
with $\Delta\bar\sigma_+$ ($\Delta\bar\sigma_-$) pointing in the signal (background) direction.
One expects $\Delta\bar\sigma_+ > \Delta\bar\sigma_-$ from Poisson statistics, 
but cases with $\Delta\bar\sigma_+ < \Delta\bar\sigma_-$ also occur frequently.
If a fluctuation below the background is seen, $\bar\sigma_{\rm obs} < 0$, then $\Delta\bar\sigma_+$ 
is used in both terms
in Eq.~(\ref{qdef}), and conversely, whenever $\bar\sigma_{\rm obs} > 1$ then only 
$\Delta\bar\sigma_-$ enters. 
The factorized form~(\ref{pmh}) is a reflection of the fact that mutual correlations between ATLAS and 
CMS, 
and between the LHC and the Tevatron, are ignored.  
This is a justifiable approximation, since in the most important regions counting rates are low 
and therefore statistical uncertainties expected to dominate. 

Consider as an example the excess at 126~GeV seen by ATLAS (Sec.~\ref{ATLAS}).
In this case, $\chi^2_{B} = 9.8$ and since the H+B hypothesis itself is not perfectly matched either, 
$\chi^2_{H+B} = 1.1$, Eq.~(\ref{qdef}) gives, 
$$2\ln Q_{\rm ATLAS}(126~\mbox{GeV}) = -8.7$$
Note, that $\sqrt{\chi^2_{B}} = 3.1$ is $0.5~\sigma$ lower than the quoted local significance of 
$3.6~\sigma$.
This can be traced to the signal cross section which introduces an additional uncertainty 
(dominated by $\sim \pm 20\%$ from the gluon-fusion production~\cite{Dittmaier:2011ti}).
While it does not affect the $p$-value for an excess over background it does enter this analysis 
which is based directly on a determination of the signal strength\footnote{Alternatively, 
one could compute $Q(M_H)$ directly from the $p$-values for H+B and B hypotheses, 
but the former have not been made available for CMS, and it is preferable to treat both LHC 
experiments in identical ways.} and thus works to reduce the significance.
Conversely, the significance for background-like outcomes is enhanced.   
Thus, $Q_{\rm LHC}$ tends to be on the conservative side in the most interesting mass region,
which compensates for the neglected correlations between ATLAS and CMS.

This completes the definition of the likelihood model used for this analysis.
The last factor in Eq.~(\ref{qdef}) is the (improper) non-informative prior density chosen such that 
the variable $\ln M_H$ has a flat prior which one can argue is the most conservative 
(least informative) one for a variable defined over the real numbers. 
The numerical significance of changing to a prior which is flat in $M_H$ itself 
does not exceed the 0.1~GeV level in the determination of $M_H$.
Before discussing the results for $p(M_H)$ in Section~\ref{results}, 
I summarize some of the individual findings, and how they enter into this analysis
(see also, the recent historical account in Ref.~\cite{Schlatter:2011ia}).

\subsection{Input Data}
\label{input}
\subsubsection{ATLAS}
\label{ATLAS}

\begin{figure}
\includegraphics[scale=0.34]{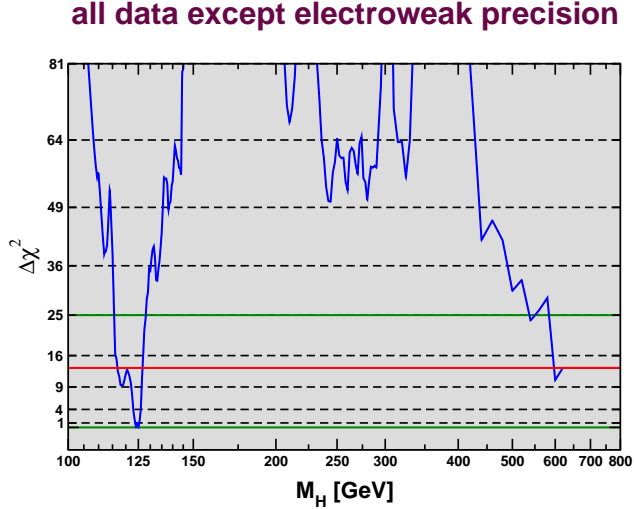}
\caption{\label{chi2noEW} 
Combination of all direct SM Higgs boson search results (see text).
Even at the $5~\sigma$ level of confidence (in the loose frequentist sense), 
there are only two remaining $M_H$ ranges (ignoring another local minimum near 540~GeV), 
namely $115.6~\mbox{GeV} < M_H < 128.1~\mbox{GeV} $ and $M_H > 584~\mbox{GeV}$.}
\end{figure}

\begin{figure}
\includegraphics[scale=0.34]{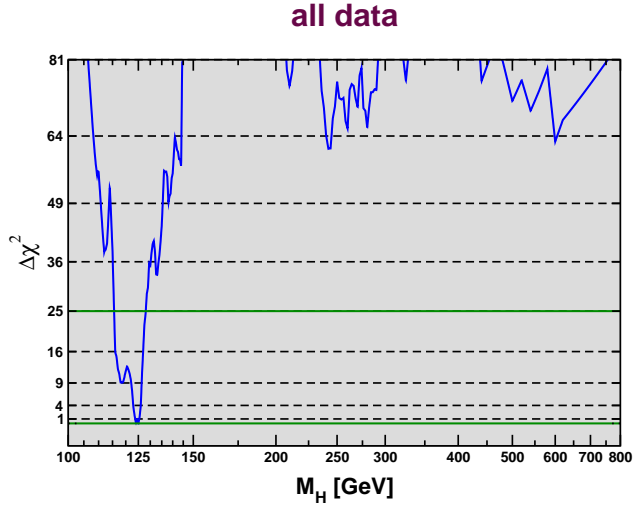}
\caption{\label{chi2all}
Combination of all direct SM Higgs boson search results with the indirect precision data.
Compared to Fig.~\ref{chi2noEW} only the low mass window remains.}
\end{figure}

ATLAS excludes the Higgs boson mass ranges from 112.7 to 115.5~GeV, from 131 to 237~GeV, 
and from 251 to 453~GeV at the 95\% CL.
An excess of events is observed for a Higgs boson mass close to $M_H =126$~GeV. 
The maximum local significance of this excess is $3.6~\sigma$ above the expected background, 
while the probability of such a fluctuation to happen anywhere in the full explored Higgs mass 
domain corresponds to a global significance of $2.3~\sigma$. 
The three most sensitive channels in this mass range,
$H \to \gamma\gamma$, $H\to ZZ^{(*)} \to \ell^+ \ell^- \ell^+ \ell^-$, and 
$H \to WW^{(*)} \to \ell^+ \nu \ell^- \bar\nu$, contribute individual local significances of 
$2.8~\sigma$, $2.1~\sigma$, and $1.4~\sigma$, respectively, to the excess~\cite{ATLAS}.

There is also an excess number of $H \to ZZ$ candidates around $M_H = 244$~GeV 
and towards the upper end of the search window (600~GeV).
They are of lower significance but describe the H+B hypothesis better than the background, given 
that 
$$2\ln Q_{\rm ATLAS}(244~\mbox{GeV})\approx 2\ln Q_{\rm ATLAS}(560~\mbox{GeV})  \approx -3$$
are negative.

\subsubsection{CMS}

Based on the $\gamma\gamma$, $b\bar{b}$, $\tau^+ \tau^-$, $W^+W^-$, and $ZZ$ decay channels,
CMS excludes the Higgs mass range from 127~GeV to the upper end of the search interval of 
600~GeV (95\% CL).
In the remaining search interval between 110 and 127~GeV two excesses are observed:
three candidate $H\to ZZ^{(*)} \to \ell^+ \ell^- \ell^+ \ell^-$ events were reconstructed consistent with 
$M_H =119.5$~GeV, compared to 1.7 (0.7) expected events for the H+B (B) hypothesis. 
While this is corroborated by an excess in $H \to WW^{(*)}$
and also in the less significant $b\bar{b}$ and $\tau^+ \tau^-$ channels,
the more sensitive $H \to \gamma\gamma$ channel shows a deficit below background. 
When combined, Eq.~(\ref{qdef}) yields,
$$2\ln Q_{\rm CMS}(119.5~\mbox{GeV}) = -5.6$$

On the other hand, there is an excess in $H \to \gamma\gamma$ corresponding to 
$M_H = 123.5$~GeV.
The signal strength is $1.7\pm 0.8$ times the expected one which amounts to a local significance of 
$2.3~\sigma$.
Including the other channels 
--- which are consistent with both the B and H+B hypotheses ---
gives a local (global) significance of 2.6 $(1.9)~\sigma$~\cite{CMS}.
When combined these data match perfectly with $M_H = 124$~GeV ($\chi^2_{H+B} = 0$), and 
Eq.~(\ref{qdef}) gives,
$$2\ln Q_{\rm CMS}(124~\mbox{GeV}) = -6.6,$$
where in this case the value of $\sqrt{\chi^2_{B}} = 2.6$ agrees exactly with the quoted local 
significance. 

\subsubsection{Tevatron}

The most recent combination from the Tevatron is based on 71 mutually exclusive final states from 
CDF and 94 from D\O.  
A small excess of data events is found in the mass range between 125~GeV and 155~GeV with
$$2\ln Q_{\rm Tevatron}(130~\mbox{GeV}) = -1.9,$$
while the region between 156~GeV and 177~GeV 
is excluded at the 95\% CL~\cite{CDFandD0:2011aa}.

\subsubsection{LEP~2}

The input from LEP~2 is unchanged with respect to Ref.~\cite{Erler:2010wa} which was the last 
analysis of the type presented here before the LHC started data taking in earnest.
At LEP~2 with energies up to $\sqrt{s} \approx 209$~GeV, the Higgs boson was searched for 
in the dominant ($\approx 74\%$) $b\bar{b}$ decay channel, produced in the Higgsstrahlung 
process, $e^+ e^- \to ZH$. 
In addition, the $H \to \tau^+ \tau^-$ channel ($\approx 7\%$) was studied for the $Z$ boson decaying 
into two jets. 
The combination~\cite{Barate:2003sz} of the four experiments, all channels and all $\sqrt{s}$ values, 
resulted in the nominal lower bound, $M_H \geq 114.4$~GeV.  
However, the combined data are neither particularly compatible with the hypothesis 
$M_H = 115$~GeV (15\% CL), nor with background only (9\% CL).
The reason is that the results by ALEPH are by themselves in very good agreement with 
$M_H \approx 114$~GeV (due to an excess in the 4-jet channel) thereby strongly rejecting 
the background only hypothesis, while the results based on the other channels and experiments 
(especially DELPHI) are incompatible with any signal.
Overall, a signal for $115~\mbox{GeV} \leq M_H \leq 119.5~\mbox{GeV}$ is favored by the data, but 
not with high significance,
$$2\ln Q_{\rm LEP}(117~\mbox{GeV}) = -1.7$$

The combination of all direct search results are illustrated in Fig.~\ref{chi2noEW}. 
Shown is the $\chi^2$ difference relative to the most signal-like Higgs mass of 125~GeV,
$$\Delta\chi^2 \equiv -2 \ln {p(M_H)\over p(125~\mbox{GeV})}, $$
where $-2\ln p(125~\mbox{GeV}) = 13.2$.
This value is indicated by the red line in the figure and corresponds to vanishing reach or else to 
cases where the overall search results are equally well (or poorly) described by the H+B and 
B hypotheses.

\subsubsection{Precision Data}

The input electroweak precision data are dominated by the results of the $Z$-pole experiments at 
LEP and the SLC~\cite{Zpole:2005ema} and correspond basically to those described in detail 
in Ref.~\cite{PDG2012}.
Despite a few discrepancies, the fit describes well the data with a $\chi^2/{\rm d.o.f.} = 45.6/42$. 
The probability of a larger $\chi^2/{\rm d.o.f.}$ is 33\%. 
Only the muon magnetic moment anomaly from BNL~\cite{Bennett:2006fi}
and the $b$-quark forward-backward asymmetry from LEP~1
are currently showing large ($3.0~\sigma$ and $2.7~\sigma$) deviations. 
In addition, the polarization asymmetry from SLD differs by $1.7~\sigma$. 
The effective $\nu$-quark coupling $g_L^2$ from NuTeV~\cite{Zeller:2001hh} is nominally in conflict 
with the SM, as well, but the precise status is under investigation by the Collaboration.

By themselves, the precision data give the $1~\sigma$ result,
$$M_H = 99^{+28}_{-23}~\mbox{GeV},$$
which covers exactly the low mass range not yet excluded by CMS.  
Compared to previous analyses~\cite{Erler:2000cr,Erler:2010wa} the indirect precision data now 
play a less pronounced
r$\hat{\rm o}$le as they do not have much discriminatory power within the remaining low mass 
window, $115.5~\mbox{GeV} < M_H <  127$~GeV, since
$$\chi^2_{\rm EW}(127~\mbox{GeV}) - \chi^2_{\rm EW}(115.5~\mbox{GeV}) = 0.63$$
However, they are the only source of information in the high mass region, $M_H \gtrsim 600$~GeV, 
which is currently beyond the reach of the LHC.  
And they are crucial to guarantee a normalizable posterior density.

\begin{figure}
\includegraphics[scale=0.34]{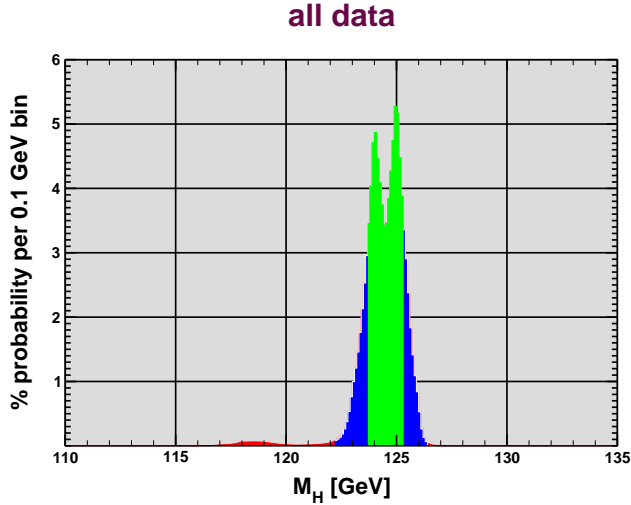}
\caption{\label{mhall}
The normalized probability distribution of $M_H$ in the low mass region based on all data.
Shown in green (blue) is the 68\% (98.2\%) CL highest probability density region.}
\end{figure}

The combination of all direct search results with the indirect precision data is shown in 
Fig.~\ref{chi2all}. 
Compared to Fig.~\ref{chi2noEW} the high mass region is now also ruled out at the $8~\sigma$ 
level\footnote{This is ignoring the fact that perturbation theory becomes unreliable for Higgs masses 
near the unitarity bound of about 800~GeV and beyond, so that the exclusion in those regions is 
rather of qualitative nature.}.

\section{Results}
\label{results}
\begin{figure}
\includegraphics[scale=0.34]{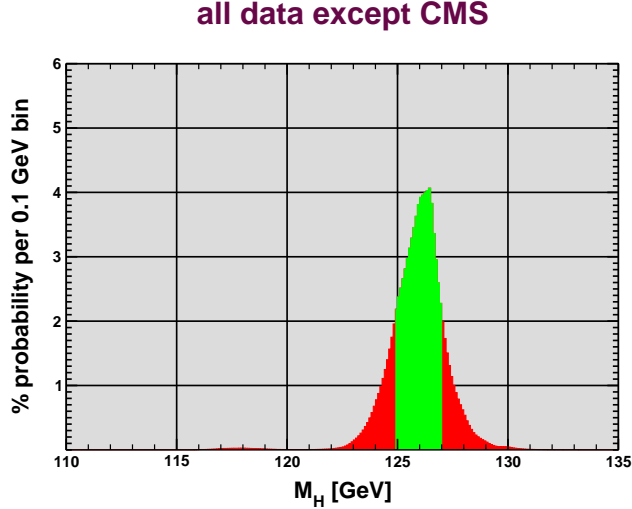}
\caption{\label{mhnoCMS}
The normalized probability distribution of $M_H$ in the low mass region based on all data except for 
CMS.
Shown in green is the 68\% CL highest probability density region, which is given by 
$M_H = 126.4^{+0.6}_{-1.5}$~GeV.}
\end{figure}

\begin{figure}
\includegraphics[scale=0.34]{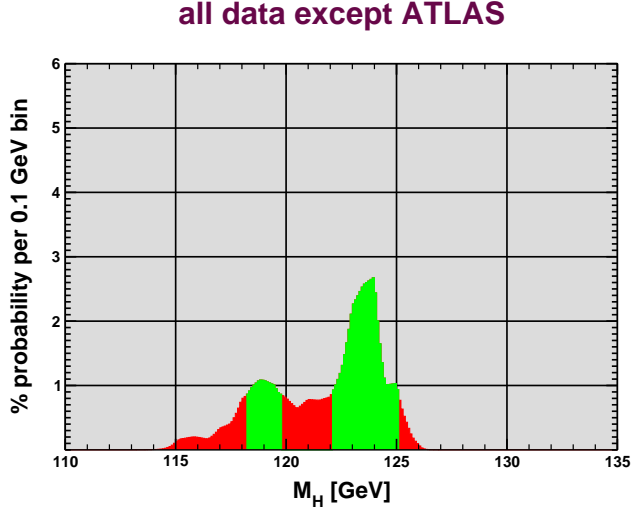}
\caption{\label{mhnoATLAS}
The normalized probability distribution of $M_H$ in the low mass region based on all data except for 
ATLAS.
Shown in green are the 68\% CL highest probability density regions, which are given by the two 
ranges, $M_H = 118.8^{+1.0}_{-0.6}$~GeV and $M_H = 123.9^{+1.2}_{-1.8}$~GeV.}
\end{figure}

\begin{figure}
\includegraphics[scale=0.34]{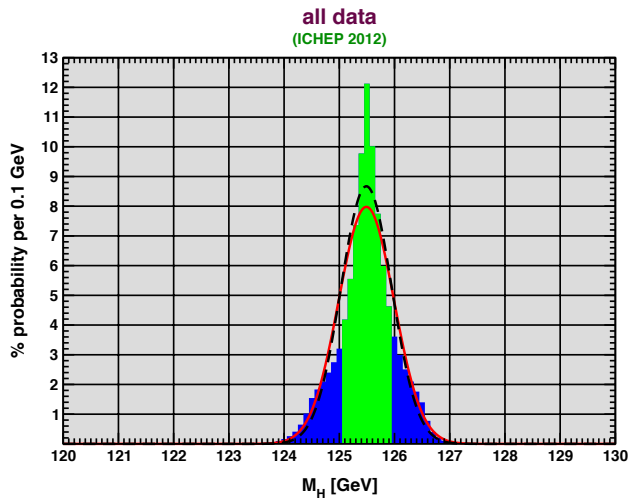}
\caption{\label{ICHEPall}
The normalized probability distribution of $M_H$ in the low mass region based on all data
available in the summer of 2012,
with the 68\% CL highest probability density region highlighted (in green).
Also shown are two reference Gaussian: 
(i) the dashed one (in black) is centered around the median, $M_H =  125.50$~GeV, 
and has the same width as the 68\%~CL {\em central\/} probability interval, $\pm 0.46$~GeV;
(ii) the solid one (in red) is based on mean and variance, $M_H = 125.49 \pm 0.50 \mbox{ GeV}$.}
\end{figure}

The main result of this communication is the normalized probability distribution of $M_H$ displayed 
in Fig.~\ref{mhall}, which is based on all available data as summarized in Sec.~\ref{input}
(see Fig~\ref{ICHEPall} below for the most recent data).
Indicated in green is the 68\% CL allowed highest probability density region.
It is given by the range,
$123.7~\mbox{GeV} \leq M_H \leq 125.3~\mbox{GeV}$,
which I write in a more colloquial form as,
\begin{equation}
M_H = 124.5 \pm 0.8~\mbox{GeV},
\label{mh}
\end{equation}
even though the central value is close to the minimum within this range rather than representing 
the mode. 
Eq.~(\ref{mh}) does contain, however, both modes at 124 and 125~GeV which originate from CMS 
and ATLAS, respectively. 
Nominally (as reviewed in Sec.~\ref{ATLAS}) the latter would be expected to be closer to 126~GeV 
(see Fig.~\ref{mhnoCMS}).
However, CMS does not see a signal there, and the peak is effectively cut, lowered, and shifted.
On the other hand, the CMS peak at 124~GeV  (see Fig.~\ref{mhnoATLAS}) 
is perfectly consistent with ATLAS, so its peak gets enhanced in the combination.
The second CMS peak at 119.5~GeV, however, is clearly disfavored by ATLAS.

Eq.~(\ref{mh}) also contains the mean of the distribution given by $M_H = 124.4 \pm 1.0$~GeV,
and is almost identical to the 68\% CL {\em central\/} interval, 
where the median happens to coincide with the central value of Eq.~(\ref{mh})

When taken together the twin peak structure in Fig.~\ref{mhall} contains the highest probability 
density at the 98.2\% CL corresponding to $2.4~\sigma$. 
This is obtained by summing the probability under each mass bin that is higher than the highest 
probability mass bin under the subleading peak near 119.5~GeV.
This may be set against the 3.6 and 2.6 $\sigma$ maximal local $p$-factors for background 
fluctuations quoted by ATLAS and CMS, respectively, or to the ``de-rated" significances of 2.2 and 
0.6~$\sigma$ after accounting for the so-called look elsewhere effect (LEE)~\cite{arXiv:1005.1891}
which is applicable if the location for a hypothetical excess is {\em a priori\/} unknown.  
The necessity for the LEE adjustment ({\em i.e.\/}, accounting for trial factors) arises from the 
frequentist set-up.
ATLAS and CMS estimate their trial factors based on observed local data fluctuations but the 
followed procedure~\cite{LLE} is bound to be somewhat arbitrary.
Among other things it depends on what one considers {\em a priori\/} excluded by previous 
experiments or data sets.
For example, if ``elsewhere" is restricted to the low mass region up to about 145~GeV,
the de-rated significances read 2.5 and $1.9~\sigma$, respectively.

It is amusing that in this way considerations of prior knowledge re-enter the frequentist framework,
which is sometimes chosen over the Bayesian one specifically to avoid prior densities\footnote{Any 
reasonable {\em probabilistic\/} model is necessarily equivalent to a Bayesian model with some prior 
even though the choice of prior may be highly implicit. 
This is so, because the proof of Bayes` theorem needs no assumptions other than the axioms of 
probability theory.}.
But as can be seen from the discussion in the previous paragraph, 
the dependence on prior knowledge of the LEE is very strong in the case of CMS,
and still $0.3~\sigma$ for ATLAS~\cite{ATLAS}.
In contrast, it is negligible for the results presented in this work.

\section{Conclusions and Outlook}
\label{outlook}
\begin{figure}
\includegraphics[scale=0.34]{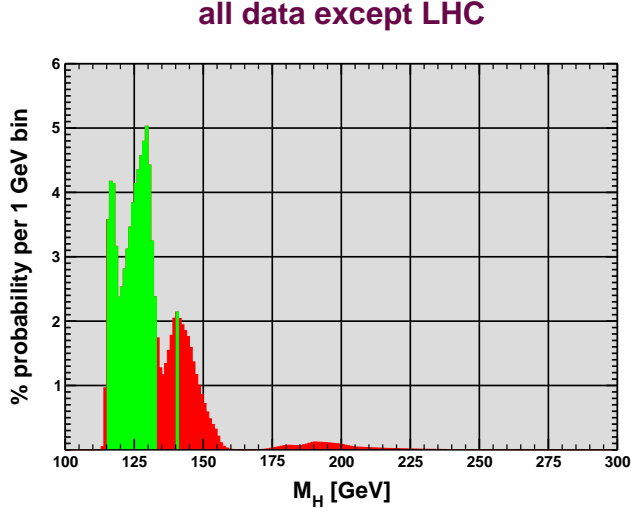}
\caption{\label{mhnoLHC}
The normalized probability distribution of $M_H$ in the low and intermediate mass regions 
based on all data except for the LHC.
Shown in green is the 68\% CL highest probability density.}
\end{figure}

I have collected all experimental information relevant to determine the Higgs boson mass,
and performed a simultaneous analysis.  
The result, $M_H = 124.5 \pm 0.8~\mbox{GeV}$, is remarkably precise and, of course, driven in large 
parts by the LHC 
(see Fig.~\ref{mhnoLHC} for the probability density when the LHC data are removed).
Incidentally, $M_H$ is determined to slightly higher absolute and slightly lower relative accuracy 
than the top quark mass~\cite{Lancaster:2011wr}, and if the SM is correct then the mass of the Higgs 
boson would have been measured accurately before its existence is indisputably confirmed.

In addition to providing a well-defined determination of $M_H$, the Bayesian statistical model 
employed here also establishes an unambiguous measure of significance.
The highest probability density under the twin peaks in Fig.~\ref{mhall} integrates to 98.2\% 
corresponding to $2.4~\sigma$.
Assuming the two LHC experiments see identical results with the next two or three data sets of the 
same size, this would increase to $4.6~\sigma$ or $5.4~\sigma$, respectively. 
Thus, the conventional $5~\sigma$ should be reached roughly with an additional 12~fb$^{-1}$ per 
experiment of data.
This can be achieved in 2012 with a luminosity corresponding to a one third increase relative to 
a good LHC week in October of 2011
(for example, by decreasing the effective beam size, $\beta^*$, by about 25\%). 

\appendix
\section{Updated notes and figures}
\label{updates}
While this article was under consideration for publication, both the 
ATLAS~\cite{Aad:2012gk} and CMS~\cite{Chatrchyan:2012gu} Collaborations
announced the observation of a new boson with mass, respectively, given by
$M = 126.0 \pm 0.4 \pm 0.4$~GeV and $M = 125.3 \pm 0.4 \pm 0.5$~GeV.
A simple weighted average would give, $M = 125.7 \pm 0.4$~GeV.
The datasets used correspond to integrated luminosities of up to approximately 5.1~fb$^{-1}$ 
collected at $\sqrt{s} = 7$~TeV in 2011 and up to 5.8~fb$^{-1}$ at $\sqrt{s} = 8$~TeV in 2012, and
the local significances are quoted at $5.9$ and $5.0~\sigma$ for the two experiments, respectively.
The CDF and D\O\ Collaborations~\cite{CDFandD0:2012zzl} also released a much improved 
analysis of their data, revealing an excess in the 115 to 140~GeV mass range with a local 
significance of $3.0~\sigma$. 
  
\begin{figure}
\includegraphics[scale=0.34]{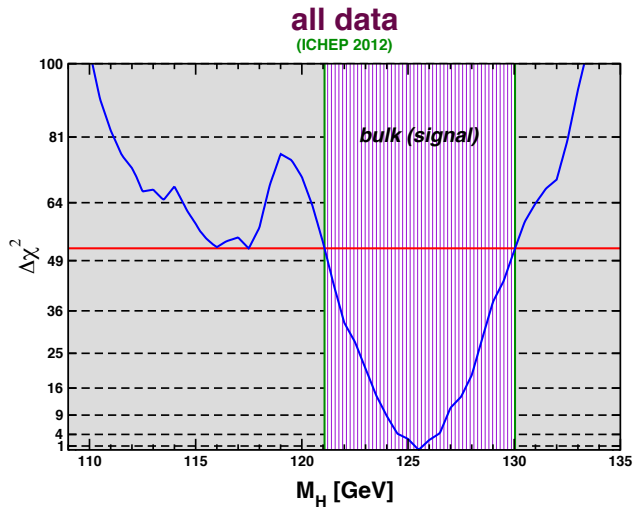}
\caption{\label{chi2_ICHEP}
Combination of all direct SM Higgs boson search results with the indirect precision data,
following the updates for the summer conferences of 2012.
This is to be compared with Fig.~\ref{chi2all}, but here the focus is on the low mass region.
It is illustrated how the bulk (signal) region can be defined unambiguously,
permitting the extraction of the total signal probability.
Using the inverse error function, this can then be translated back into the number of standard
deviations of the signal without reference to the LEE.}
\end{figure}

Here I update the results reflecting these developments.
The 68\%~CL allowed highest probability density range is now 
$125.02 \mbox{ GeV} \le M_H \le 125.95$~GeV, or in short, 
\begin{equation}\label{mhICHEP}
M_H = 125.5 \pm 0.5 \mbox{ GeV}.
\end{equation}
Unlike the remarks following Eq.~(\ref{mh}), the latest data combine to a nearly
bell-shaped curve as shown in Fig.~\ref{ICHEPall}, with coinciding mean, median, and mode.
The significance of the bulk region of the probability distribution,
$121.1 \mbox{ GeV} \le M_H \le 130.0$~GeV
--- according to the method introduced here and illustrated in Fig.~\ref{chi2_ICHEP} --- 
is $6.8~\sigma$, {\em i.e.}, the tail regions contain a probability of $9\times 10^{-12}$.
Given the greater effectiveness of the 8~TeV LHC for Higgs searches, 
this is is consistent with but exceeds somewhat the expectation expressed in Sec.~\ref{outlook}.

\begin{figure}
\includegraphics[scale=0.34]{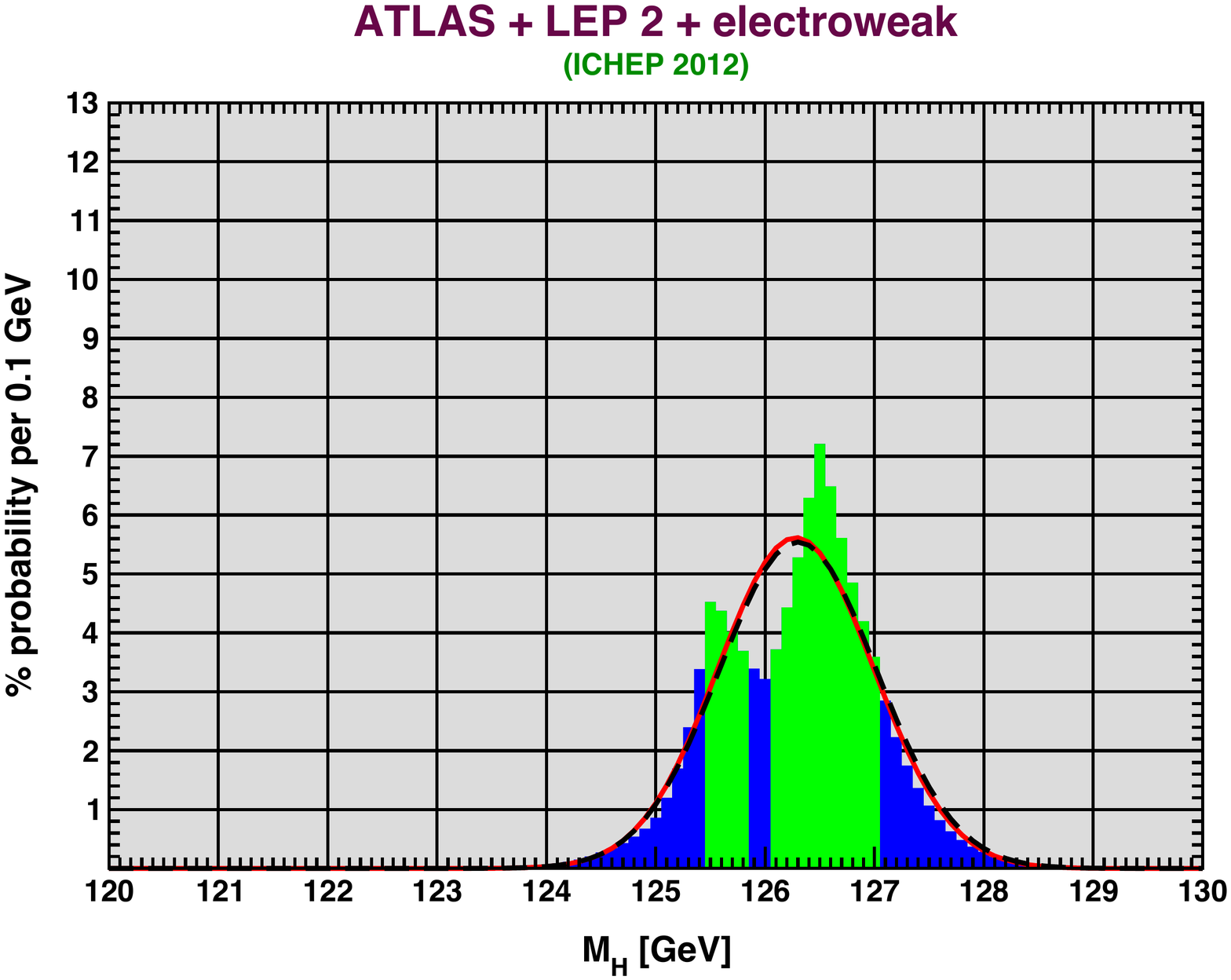}
\caption{\label{ICHEPATLAS}
The normalized probability distribution of $M_H$ in the low mass region based on Higgs search 
results from LEP~2 and ATLAS, as well as the electroweak precision data,
with the 68\%~CL highest probability density region highlighted (in green).
Also shown are the median (in black) and mean (in red) motivated reference Gaussian densities
({cf.}\ Fig.~\ref{ICHEPall}), 
with $M_H = 126.30 \pm 0.72$~GeV and $M_H = 126.28 \pm 0.71$~GeV, respectively,
which happen to be almost identical in this case.}
\end{figure}

\begin{figure}
\includegraphics[scale=0.34]{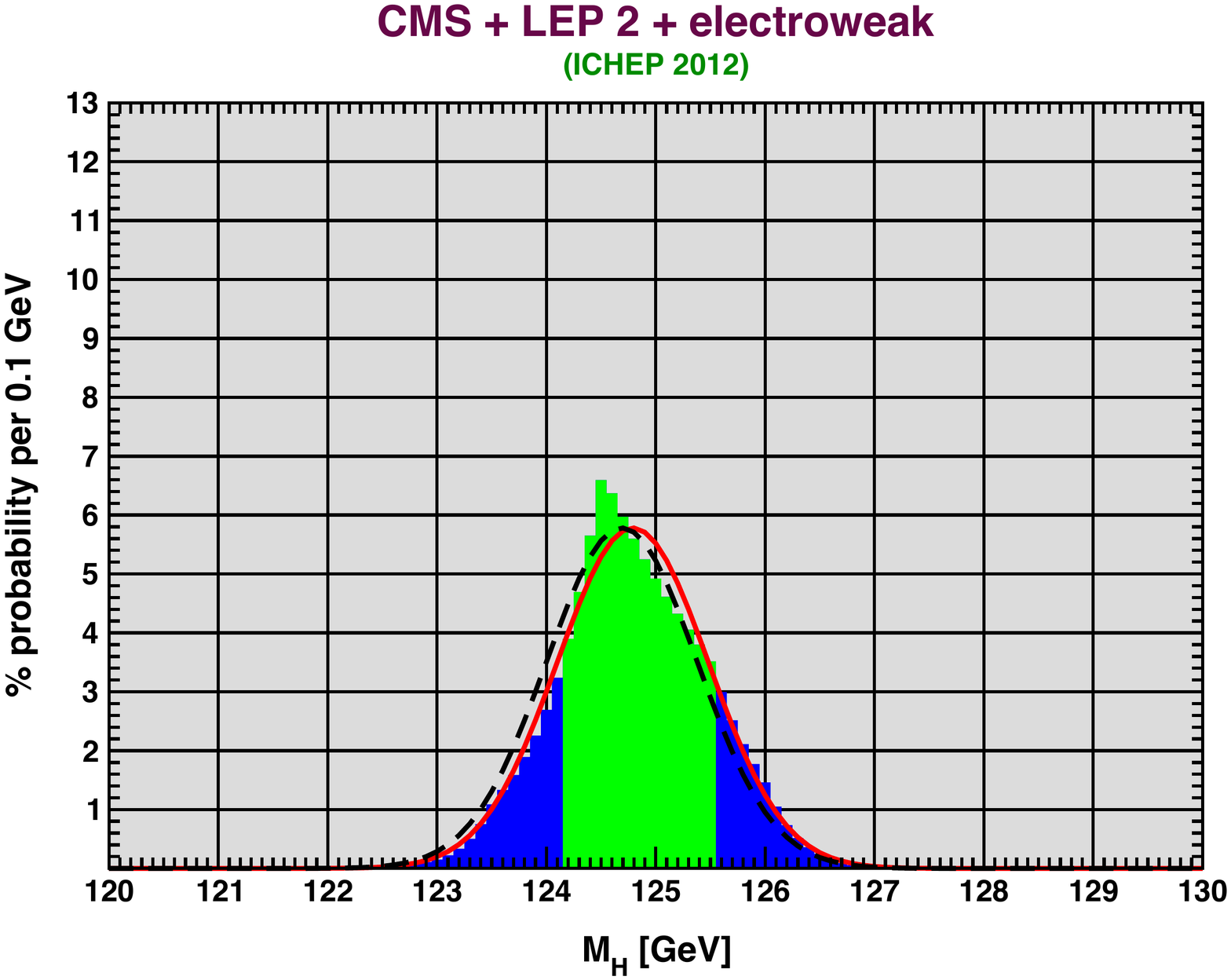}
\caption{\label{ICHEPCMS}
The normalized probability distribution of $M_H$ in the low mass region based on Higgs search 
results from LEP~2 and CMS, as well as the electroweak precision data,
with the 68\%~CL highest probability density region highlighted (in green).
Also shown are the median (in black) and mean (in red) motivated reference Gaussian densities
({cf.}\ Fig.~\ref{ICHEPall}),
with $M_H = 124.69 \pm 0.69$~GeV and $M_H = 124.79 \pm 0.69$~GeV, respectively.}
\end{figure}

Similarly, Fig.~\ref{ICHEPATLAS} (Fig.~\ref{ICHEPCMS}) shows the corresponding distribution
for the combination of the ATLAS (CMS) and LEP~2 Higgs search results with the 
electroweak precision data. The significance of the ATLAS bulk region is $4.9~\sigma$
and larger that the one from CMS ($4.2~\sigma$), but the CMS data are slightly sharper peaked
as can be seen from the reference Gaussian densities also shown in the figures.
The probability that the true $M_H$ resides in one of the tails is close to $10^{-6}$ for ATLAS and 
$3\times 10^{-5}$ for CMS.
These numbers are several orders of magnitude larger than the background 
fluctuation probabilities ($p$-values) of $1.7\times 10^{-9}$ and $3\times 10^{-7}$, respectively. 

\begin{figure}
\includegraphics[scale=0.34]{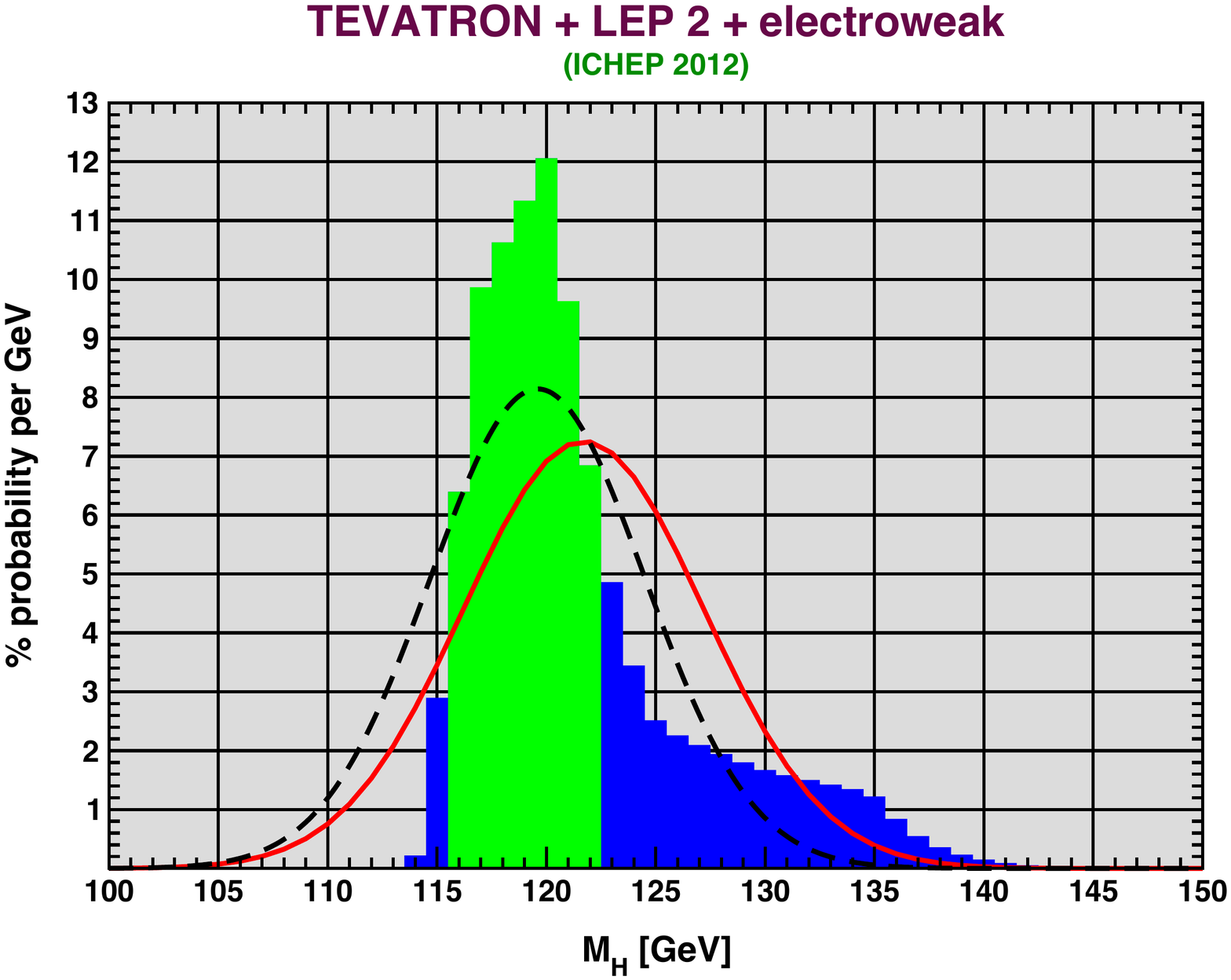}
\caption{\label{ICHEPTEVATRON}
The normalized probability distribution of $M_H$ in the low mass region based on Higgs search 
results from LEP~2 and the Tevatron, as well as the electroweak precision data,
with the 68\%~CL highest probability density region highlighted (in green).
Also shown are the median (in black) and mean (in red) motivated reference Gaussian densities,
with $M_H = 119.6 \pm 5.1$~GeV and $M_H = 121.7 \pm 5.5$~GeV, respectively
({cf.}\ Fig.~\ref{ICHEPall}).}
\end{figure}

Finally,  Fig.~\ref{ICHEPTEVATRON} shows the case for the combination of the CDF, D\O\ and 
LEP~2 Higgs search results with the electroweak precision data, with a bulk region significance 
of $3.4~\sigma$.  
The tail probability of $6 \times 10^{-4}$ is in this case {\em lower\/}
than the background $p$-value of $1.5 \times 10^{-3}$.

\begin{acknowledgments}
It is a pleasure to thank Nima Arkani-Hamed, Paul Langacker, and Edward Witten for
stimulating discussions and encouragement.
This work was supported by CONACyT (M\'exico) project 82291--F 
and by PASPA (DGAPA--UNAM).
\end{acknowledgments}

\end{document}